# A tutorial on group effective connectivity analysis, part 2: second level analysis with PEB

Peter Zeidman*[1], Amirhossein Jafarian[1], Mohamed L. Seghier[2], Vladimir Litvak[1], Hayriye Cagnan[3], Cathy J. Price[1], Karl J. Friston[1]

[1] Wellcome Centre for Human Neuroimaging
12 Queen Square
London
WC1N 3AR

[2] Cognitive Neuroimaging Unit,
ECAE,
Abu Dhabi,
UAE

[3] Nuffield Department of Clinical Neurosciences
Level 6
West Wing
John Radcliffe Hospital
Oxford
OX3 9DU

* Corresponding author

**Abstract**

This tutorial provides a worked example of using Dynamic Causal Modelling (DCM) and Parametric Empirical Bayes (PEB) to characterise inter-subject variability in neural circuitry (effective connectivity). This involves specifying a hierarchical model with two or more levels. At the first level, state space models (DCMs) are used to infer the effective connectivity that best explains a subject's neuroimaging timeseries (e.g. fMRI, MEG, EEG). Subject-specific connectivity parameters are then taken to the group level, where they are modelled using a General Linear Model (GLM) that partitions between-subject variability into designed effects and additive random effects. The ensuing (Bayesian) hierarchical model conveys both the estimated connection strengths and their uncertainty (i.e., posterior covariance) from the subject to the group level; enabling hypotheses to be tested about the commonalities and differences across subjects. This approach can also finesse parameter estimation at the subject level, by using the group-level parameters as empirical priors. We walk through this approach in detail, using data from a published fMRI experiment that characterised individual differences in hemispheric lateralization in a semantic processing task. The preliminary subject specific DCM analysis is covered in detail in a companion paper. This tutorial is accompanied by the example dataset and step-by-step instructions to reproduce the analyses.

# Contents

## 1 Introduction

Neuroimaging studies typically have a hierarchical form. At the first (within subject) level, the neural responses of individual subjects are inferred from measurements (e.g. fMRI, EEG, MEG) by specifying and fitting suitable models. The ensuing model parameters are then taken to the second (between subject) level, where the commonalities and differences across subjects are assessed. There may be further levels of the hierarchy; for example, each group of subjects may have been sampled from different populations. This naturally suggests the use of a hierarchical model that links individual subjects to the population(s) from which they were sampled. In this paper, we address the practicalities of hierarchical modelling in brain connectivity studies.

Dynamic Causal Modelling (DCM) is the predominant analysis framework for inferring *effective* connectivity; namely, the directed causal influences among neural populations that mediate an *effect* of one population on another (Friston et al., 2003). In this context, a *model* is a set of differential equations that describe the transformations from experimental stimulation through neural activity to observed data. The model has parameters, such as the strength of particular neural connections, which are estimated from the data using a Bayesian (probabilistic) method called Variational Laplace (Friston et al., 2007). This provides a probability density over the possible values of the parameters (e.g., connection strengths), as well as a score for the quality of the model, called the log-evidence or (negative) variational free energy.

Having inferred every subject's connectivity strengths, the next challenge is how to quantify the commonalities and differences across subjects. For example, one may wish to test for differences between a patient group and a control group, or investigate whether the dose of a drug alters certain connections, or whether there is a relationship between connection strengths and behavioural measures. To enable these kinds of hypotheses to be tested efficiently, DCM was recently supplemented with a hierarchical model over parameters, using the Parametric Empirical Bayes (PEB) framework (Friston et al., 2016). Readers familiar with mass-univariate analysis in neuroimaging (Statistical Parametric Mapping,

SPM) will be accustomed to the 'summary statistic' approach, which begins with quantifying effects at the first or within-subject level, followed by a second level analysis to test whether these effects are conserved over subjects. The PEB framework enables a similar workflow for DCM, illustrated schematically in Figure 1 in the context of the example fMRI experiment presented here. A DCM is specified for each subject and DCM parameters are estimated from the data (Figure 1, bottom). The parameters of interest are then collated and modelled at the second level using a General Linear Model (GLM), and any unexplained between-subject variability is captured by a covariance component model. Therefore, individual differences in connection strengths are decomposed into hypothesised group-level effects, plus any unexplained random effects (RFX), illustrated in Figure 1 (top). Having estimated the group-level parameters (e.g., group-average connection strengths), hypotheses can be tested by comparing the evidence for different mixtures of these parameters – a process referred to as Bayesian model comparison. In this tutorial, we only introduce Bayesian model comparison at the point at which a hypothesis is tested. In group studies, the hypothesis is, by definition, about between subject effects (as opposed to which DCM best accounts for subject specific data). Our focus is therefore on comparing different models of between-subject effects on (condition specific changes in) within-subject connectivity.

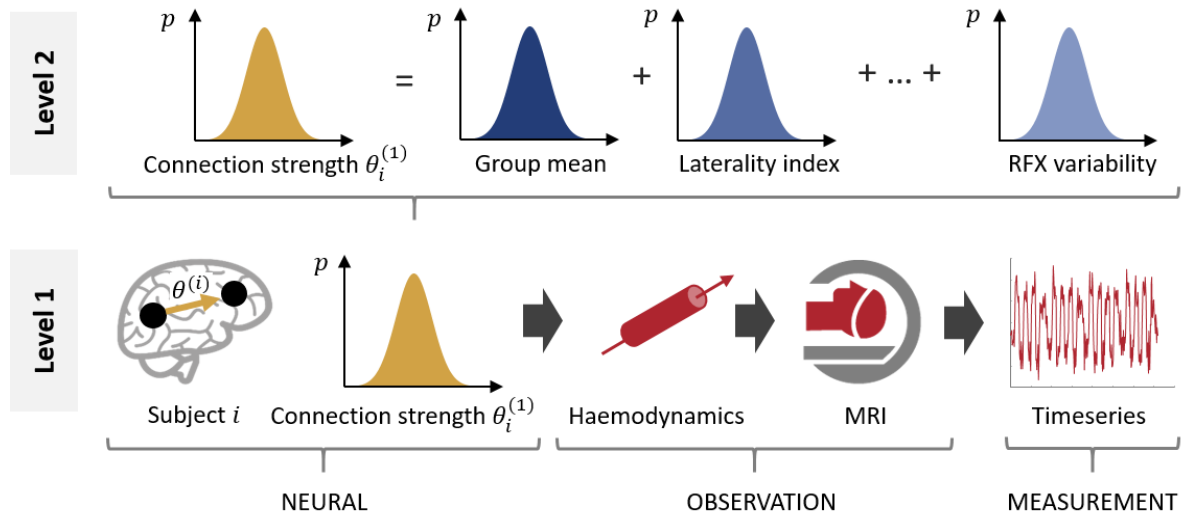

**Figure 1. The two-level Parameter Empirical Bayes (PEB) model used in our analysis.** At the first level (bottom of the figure), a forward model (DCM) describes how neural activity causes the fMRI timeseries of each subject $i = 1 \ldots N$. The parameters from the neural part of the model are taken to the group level (upper part of the figure), where they are modelled as a linear combination of a group mean, a group difference due to laterality index, covariates of no interest (not shown) and random effects (RFX) variability. Image credits: "Brain image" by parkjisun and "CT Scan" by Vectors Market from the Noun Project.

A simple alternative to using parametric empirical Bayes is to take the estimated connectivity parameters from all subjects and run classical statistics such as ANCOVA or MANCOVA. However, the PEB framework offers a number of advantages over this. As illustrated in Figure 1, estimated parameters in DCM are probability densities. Taking just the expected values (peaks) from the probability density would ignore the estimated covariance (uncertainty) of the parameters. By contrast, the PEB framework takes both the expected values and the covariance of the parameters to the group level. A useful implication is that more precise parameter estimates will have greater influence on the group-level result, so subjects with noisy data and uncertain parameter estimates will be down-weighted. Additionally, each level of the hierarchy places constraints on the level below, expressing the prior belief that it would be surprising if any individual subject had radically different connectivity than the rest of the group, beyond those sources of between-subject variability modelled explicitly. This can give improved parameter estimates at the single subject level, which could be important in settings where precise inferences about individual subjects are needed (e.g., when identifying potential biomarkers).

The PEB framework may be contrasted with an established approach for group-level modelling of effective connectivity; namely, Random Effects Bayesian Model Selection (RFX BMS), introduced in Stephan et al. (2009). With RFX BMS, multiple DCMs are specified per subject, where each DCM corresponds to one hypothesis about the network architecture. These DCMs are fitted to each subject's data. A hierarchical model is then defined, which estimates the relative probability that any randomly selected person from the population would have had their data generated by each DCM. The 'random effect' in this context is which model generated each subject's data; i.e., some subjects may have had their data generated by model 1, others by model 2, *etc*. This sort of random effect on models may be particularly appropriate when subjects vary in their neural architecture in terms of the presence or absence of particular connections; for example, when testing for functional dysconnections in schizophrenia (Friston, 1998). The RFX BMS method has also benefited from extensive validation (Rigoux et al., 2014).

The PEB approach differs fundamentally, in that is a hierarchical model with random effects on *parameters* rather than *models*. In other words, it assumes that all subjects have the same basic architecture but differ in terms of the strengths of connections within that architecture. Only one DCM is specified per subject and the estimated parameters are taken to the group level, where hypotheses about these between subject differences are tested. With this approach, parameters may be treated as 'fixed effects' or 'random effects' (i.e., sampled from a wider population) by setting appropriate priors on between-subject variability in the hierarchical model. Unlike the BMS RFX approach, hypothesised between-subject effects can be continuous behavioural or clinical measures (e.g., a brain lateralisation index – LI – in this study), rather than the presence or absence of connections. Also, by comparing the free energy of group-level PEB models, hypotheses can be tested in terms of different mixtures of first level effects (connections) and second level effects (covariates), affording a flexibility in the questions that can be addressed. Crucially, by only inverting one 'full' DCM per subject in the PEB approach – rather than multiple DCMs per subject – one does not run into the problem of different DCMs falling into different local optima – a particular problem for neural mass models used with electrophysiological data.

To illustrate PEB in this setting, we analysed data from a previously published fMRI study on the lateralisation of semantic processing, using DCM and PEB. The first level DCM analysis is described in the first part of the tutorial (please see companion paper), which focused on the specifics of fMRI data analysis. Here, we address the more generic issue of how to model commonalities and differences among subjects in effective connectivity at the group level, regardless of the imaging modality. After introducing the example dataset, we describe the specification of a PEB model and demonstrate testing hypotheses using Bayesian Model Reduction (BMR) – a particularly efficient form of Bayesian model selection. We conclude by illustrating how predictive validity can be assessed using cross-validation. The example dataset and a step-by-step guide to running these analyses are available from https://github.com/pzeidman/dcm-peb-example .

## 2 Notation

Vectors are denoted by lower case letters in bold italics ($\boldsymbol{a}$) and matrices by upper case letters in bold italics ($\boldsymbol{A}$). Other variables and function names are written in plain italics ($f$). The dot symbol ($\cdot$) on its own means multiplication and when positioned above a variable (e.g. $\dot{z}$) denotes the derivative of a variable with respect to time. An element in row $m$ and column $n$ of matrix $\boldsymbol{A}$ is denoted by $\boldsymbol{A}_{mn}$. All variables and their dimensions are listed in Table 1. To help associate methods with their implementation in the SPM software (http://www.fil.ion.ucl.ac.uk/spm/software/), MATLAB function names are provided in bold text, such as (**spm_dcm_fit.m**).

## 3 Experimental design

We used data from a previously published fMRI experiment on language lateralisation with 60 subjects (Seghier et al., 2011). To recap, this experiment investigated how the left and right frontal lobes interact during semantic (relative to perceptual) processing of words. While language is typically thought to be left lateralised, the right hemisphere also responds in language tasks, and this experiment focussed on individual differences in the degree of lateralisation. At the within-subject level, it was a balanced 2x2 factorial design with factors: stimulus type (*Words* or *Pictures*) and task (*Semantic* or *Perceptual* matching). There was also a single factor at the between-subject level: Laterality Index (LI). This is a measure of functional brain asymmetry, derived from the number of activated voxels in each hemisphere in an initial SPM analysis. A positive LI (towards +1) indicated left hemisphere dominance, whereas a negative LI (towards -1) indicated right hemisphere dominance. The main question addressed by this experiment was: what neural circuitry underlies individual differences in LI?

Analysing a group connectivity study begins by identifying which effects would be expected to occur at the within-subject level, and which would be expected at the between-subject level. Typically, within-subject effects consist of one measurement per trial, whereas between-subject effects consist of one measurement per subject. Here, the effects of words and pictures were at the within-subject level, so we included these in each subject's first level General Linear Model (GLM) and subsequent DCM, whereas LI was a between-subject measure, so we only introduced it in the second level PEB analysis. Other factors of no interest (e.g., age, gender, and handedness) were also included at the between-subjects level.

As detailed in the next section, fitting DCMs to the fMRI data provided us with estimates of parameters of interest per subject: the effects of words and pictures on the inhibitory self-connections in: 1) left ventral frontal, lvF, 2) left dorsal frontal, ldF, 3) right ventral frontal, rvF and 4) right dorsal frontal, rdF cortex. Here, we asked which mixtures of these eight parameters best explained the commonalities across subjects, and which best explained individual differences due to LI. We hypothesised that the commonalities across subjects, and the LI differences, could be expressed in:

1. the network's response to pictures and / or words,
2. the network's response in dorsal and / or ventral regions,
3. the network's response in left and / or right hemisphere regions.

These hypotheses formed independent *factors* in our analysis, and we will detail the specification and comparison of group level models that varied across these factors (Figure 2). We will also illustrate the comparison of group level models in a less constrained manner, using an automatic search of the hypothesis or model space. First, we will briefly reprise the first level DCM analysis, which quantified the effects of pictures and words on each connection for each subject.

**Factor 1**: Modulation of regions by (P)ictures or (W)ords?

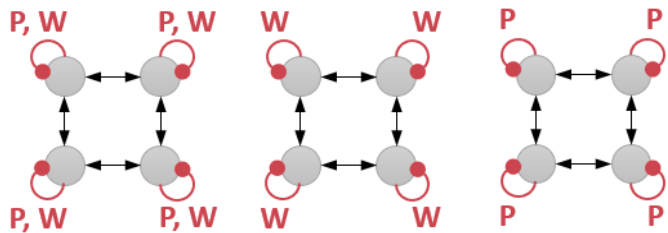

**Factor 2**: Modulation of dorsal or ventral regions?

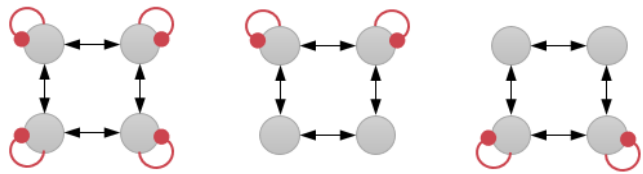

**Factor 3**: Modulation of left or right hemisphere regions?

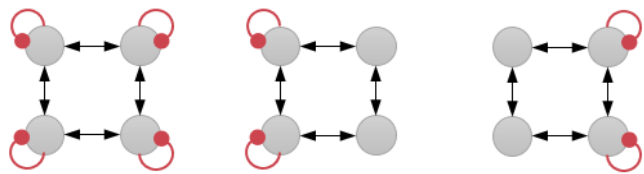

**Figure 2 Model space**. Each picture is a network of four frontal brain regions. The grey circles in the top row of each picture are left dorsal (ldF) and right dorsal (rdF) frontal cortex, and the circles in the bottom row of each picture are left ventral (lvF) and right ventral (rvF) frontal cortex. The red curved lines indicate inhibitory self-connections that were allowed to be modulated by the task. Models varied according to the three factors shown. Additionally, a 'null' model (not shown) was specified with no modulation to serve as a baseline. This gave rise to $3 \times 3 \times 3 + 1 = 28$ models.

## 4   First level analysis: DCM

Our connectivity analysis began by specifying one DCM for fMRI model for each subject (Figure 1, bottom). Here, we focussed on the subset of parameters from these models that quantified the change in each brain region's inhibitory self-connection due to words and pictures (matrix $B$ in the DCM neural model). These parameters had values greater than zero if there was an increase in self-inhibition due to the stimuli, and so a reduction in sensitivity to inputs from the rest of the network. They had values less than zero if there was a reduction in self-inhibition (i.e., disinhibition) due to the stimuli, thus increasing sensitivity to inputs from the network. We shall write these parameters of interest as $\boldsymbol{\theta}_i^{(1)}$ for subject $i$, where the superscript number 1 indicates the first level of analysis.

We then inverted each subject's DCM. This involves finding the (posterior density over) parameters that maximises a score for the model, called (negative) variational free energy $F^{(1)}$. This approximates (a lower bound on) the log model evidence $\ln p(Y_i|m)$, which is the log of the probability of having observed the data $Y_i$ given the model $m$. In short, Bayesian model inversion provides an estimate of log evidence and a multivariate normal probability density over the parameters:

$$p(\boldsymbol{\theta}_i^{(1)}|Y_i) \sim N(\boldsymbol{\mu}_i, \boldsymbol{\Sigma}_i) \qquad 1.$$

where $\boldsymbol{\mu}_i$ is a vector of the expected values of the parameters (e.g., the connection strengths) and $\boldsymbol{\Sigma}_i$ is their posterior covariance matrix. Elements on the leading diagonal of $\boldsymbol{\Sigma}_i$ are the posterior variance (uncertainty) of each parameter and the off-diagonal elements are their corresponding covariance.

The estimation scheme, called variational Laplace (Friston et al., 2007), tunes the parameters (e.g. connection strengths) with the objective of making the predicted neuroimaging timeseries match the observed timeseries as closely as possible (maximising accuracy), while penalizing movement of the parameters away from their prior values (minimizing complexity). The ideal compromise between accuracy and complexity maximises the free energy $F^{(1)}$. However, this process inherently involves uncertainty: multiple settings of the parameters could provide similarly good explanations for the data. Furthermore, there will typically be some collinearity between the parameters, meaning their effect on the predicted timeseries cannot be disambiguated. DCM quantifies this uncertainty by returning a full probability density over the parameters – both their expected values and covariance - as defined in Equation 1.

Having inverted each subject's DCM, we ran a PEB analysis (**spm_dcm_peb.m**). The first step is to collate the parameter estimates of interest from all subjects $\boldsymbol{\theta}^{(1)}$. There were two quantities from each subject: the expected values the parameters $\boldsymbol{\mu}^{(1)} = \left[\boldsymbol{\mu}_1^{(1)}, \dots, \boldsymbol{\mu}_N^{(1)}\right]^T$ and their covariance matrices $\boldsymbol{\Sigma}^{(1)} = (\boldsymbol{\Sigma}_1^{(1)}, \dots, \boldsymbol{\Sigma}_N^{(1)})$. By only taking parameters to the group level that are needed to test hypotheses, we can ensure an efficient testing of model's or hypotheses about group effects. For example, a priori, we expect laterality to effect differences in neuronal coupling – but not haemodynamics. This means evidence for an effect of lateralisation from neuronal parameters may be confounded by including haemodynamic parameters; especially in the presence of collinearity or conditional dependence. Effectively, parameters not included in the group level analysis (e.g. related to haemodynamics), are treated as *fixed effects* – meaning they are not treated as if they are sampled from a wider population – and are therefore released from the constraint of being concentrated around some group mean.

# 5 Second level analysis: Parametric Empirical Bayes (PEB)

## 5.1 Theory

In this section, we offer a brief overview of the theory behind the PEB framework, before illustrating its application to the example dataset. The framework specifies a hierarchical statistical model of connectivity parameters:

$$\boldsymbol{\theta}^{(2)} = \boldsymbol{\eta} + \boldsymbol{\epsilon}^{(3)}$$

$$\boldsymbol{\theta}^{(1)} = \mathbf{X}\boldsymbol{\theta}^{(2)} + \boldsymbol{\epsilon}^{(2)} \qquad 2.$$

$$\mathbf{y}_i = \Gamma_i(\boldsymbol{\theta}_i^{(1)}) + \mathbf{X_0}\boldsymbol{\beta_i} + \boldsymbol{\epsilon}_i^{(1)}$$

Starting with the last line of Equation 2, the observed neuroimaging data $\mathbf{y}_i$ for subject $i$ are modelled as having been generated by a DCM (or by any nonlinear model) – denoted by $\Gamma_i$ – with parameters $\boldsymbol{\theta}_i^{(1)}$. Any known uninteresting effects, such as the mean of the signal, are modelled by a GLM with design matrix $\mathbf{X_0}$ and parameters $\boldsymbol{\beta_i}$. The observation noise is modelled as residuals $\boldsymbol{\epsilon}_i^{(1)}$. The second line of Equation 2 is the second level of the PEB model. It says that the vector of all subjects' neural parameters $\boldsymbol{\theta}^{(1)}$, ordered by subject and then by parameter, can be described by a GLM with design matrix $X$ and group-level parameters $\boldsymbol{\theta}^{(2)}$.

The group-level design matrix $X$ encodes the hypothesised sources of variation across subjects, with one column for each experimental effect (called regressors or covariates). For every column in the design matrix there is a corresponding entry in parameter vector $\boldsymbol{\theta}^{(2)}$ which is estimated from the data. Each of these parameters is the group-level effect of one covariate (e.g. LI, handedness or gender) on one connection. Any differences between subjects not captured by this model are defined as zero-mean (I.I.D.) additive noise $\boldsymbol{\epsilon}^{(2)}$. These are the *random effects* (RFX). In other words, from the point of view of a generative model, to generate a single subject's data, one would first sample a parameter vector from the prior distribution (level 3), and add a *random effect* for this subject (level 2). Finally, one would generate data using the DCM and add some observation noise (level 1). More levels could be added to the hierarchy; for example, to model nested groups of subjects. However, Equation 2 stops at two levels and specifies that parameters $\boldsymbol{\theta}^{(2)}$ would have fixed prior expected value $\boldsymbol{\eta}$ and residuals $\boldsymbol{\epsilon}^{(3)}$, expressed in the first line of Equation 2. In summary, this hierarchical model partitions the variability in connectivity parameters across subjects into hypothesized group-level effects $X\boldsymbol{\theta}^{(2)}$ and uninteresting between-subject variability $\boldsymbol{\epsilon}^{(2)}$.

To turn this into a statistical (i.e. probabilistic) model, we first define the probability density over the error terms:

$$\epsilon_i^{(1)} \sim N(\mathbf{0}, \Sigma_i^{(1)})$$

$$\epsilon^{(2)} \sim N(\mathbf{0}, \Sigma^{(2)}) \quad \quad 3.$$

$$\epsilon^{(3)} \sim N(\mathbf{0}, \Sigma^{(3)})$$

Where the covariance matrices $\Sigma_i^{(1)}$, $\Sigma^{(2)}$ and $\Sigma^{(3)}$ are I.I.D. and define the zero-mean additive noise at each level of the hierarchy. The model can then be written in terms of probability densities:

$$P(\mathbf{Y}, \boldsymbol{\theta}^{(1)}, \boldsymbol{\theta}^{(2)}) = \sum_i \underbrace{\ln p(\mathbf{y}_i|\boldsymbol{\theta}^{(1)})}_{\text{1st level}} + \underbrace{\ln p(\boldsymbol{\theta}^{(1)}|\boldsymbol{\theta}^{(2)})}_{\text{2nd level}} + \underbrace{\ln p(\boldsymbol{\theta}^{(2)})}_{\text{Group priors}}$$

$$p(\boldsymbol{\theta}^{(2)}) = N(\boldsymbol{\eta}, \Sigma^{(3)}) \quad \quad 4.$$

$$p(\boldsymbol{\theta}^{(1)}|\boldsymbol{\theta}^{(2)}) = N(\mathbf{X}\boldsymbol{\theta}^{(2)}, \Sigma^{(2)})$$

$$p(\mathbf{y}_i|\boldsymbol{\theta}_i^{(1)}) = N(\Gamma_i(\boldsymbol{\theta}_i^{(1)}), \Sigma_i^{(1)})$$

Equation 4 specifies the joint probability of all three quantities of interest; namely, data from all subjects $Y$, the DCM parameters of all subjects $\boldsymbol{\theta}^{(1)}$ and the group-level parameters $\boldsymbol{\theta}^{(2)}$. The last equality states that the timeseries for subject $i$ is generated by their DCM, with covariance (uncertainty) determined by their observation noise. The penultimate equality links the neural parameters of the individual subjects to group-level effects, and the second equality defines the priors on the group-level parameters. Each level acts as a constraint on the level below – meaning that group level parameters constrain the estimates of the individual subjects.

Configuring the PEB model for an experiment only requires specifying the second level design matrix $X$, detailed in the next section. The software provides default prior probability densities for the between-subject variability $\Sigma^{(2)}$ and the second level neural parameters $\boldsymbol{\theta}^{(2)}$. In brief, the priors used for inverting subject-specific DCMs are used as the prior variance around the group mean, while random effects at the between subject level are set to 1/16 of this variance. This assumes that people only acquire data from multiple subjects when the random effects from subject to subject are less than (i.e., have a quarter of the standard deviation) the range of values that one expects *a priori*. For details, please see *Appendix 1: PEB connectivity priors specification* and *Appendix 2: PEB random effects specification*.

## 5.2 PEB: Design matrix specification

The group-level design matrix $X$ defines the hypotheses about between-subject variability. It is specified in two parts: between-subject effects $X_B$ and within-subject effects $X_W$. The between-subject design matrix (Figure 3, left) encodes covariates or regressors in each column, with one row per subject. The software implementation in SPM expects the first column to be ones (to model the commonalities across subjects; i.e., constant or group mean) and the second column is usually the effect of interest. Other covariates are placed in the subsequent columns. For this experiment, we included covariates encoding the group mean, LI score, handedness, gender and age. (We computed the LI score by taking the first principal component of the four different measures used in Seghier et al. (2011).)

An important decision – when preparing the between-subjects design matrix – is whether to mean-centre regressors which follow the (first) constant term. If they are mean-centred, then the first regressor (the column of ones) would correspond to the mean (changes in) connectivity over subjects, and the between-subject effects would add to or subtract from this. If they were not mean-centred, then the first regressor would correspond to the baseline or intercept of the model. Here, we mean-centred, endowing the first regressor with the interpretation of the group mean connectivity.

The within-subjects design matrix (Figure 3, middle) defines which DCM connectivity parameters can receive between-subject effects (via setting each element on the leading diagonal to 1 or 0). We wanted all between-subjects effects to have the potential to influence all DCM parameters of interest, so we set this to be the identity matrix (ones on the leading diagonal, zeros elsewhere). Here, this corresponds to the effects of pictures on each of the four regions, followed by the effects of words on each of the four regions. The two parts of the design matrix are combined by the software (**spm_dcm_peb.m**) according to:

$$X = X_B \otimes X_W \qquad 5.$$

Where $\otimes$ is the Kronecker tensor product, which replicates matrix $X_W$ for each element of $X_B$. This is a subtle aspect of PEB that makes things slightly more complicated than simply modelling experimental effects on some response variable. Here, we have to specify what the experimental (between-subject) effects actually acts upon; namely, which (within-subject) parameters. For example, the effect of laterality may be expressed in all connectivity parameters – or just one parameter. By default, the software assumes that every experimental (between-subject) effect can be expressed on every (within-subject) parameter.

A part of the resulting design matrix $X$ is shown in Figure 3 (right). The first eight columns encode the group mean of each connectivity parameter. Columns 9-16 encode the effect of LI on each connectivity parameter, and the remaining columns encode the effects of handedness, gender and age. The rows are ordered subject-wise, corresponding to DCM parameters of interest from subject one, then from subject two, etc.

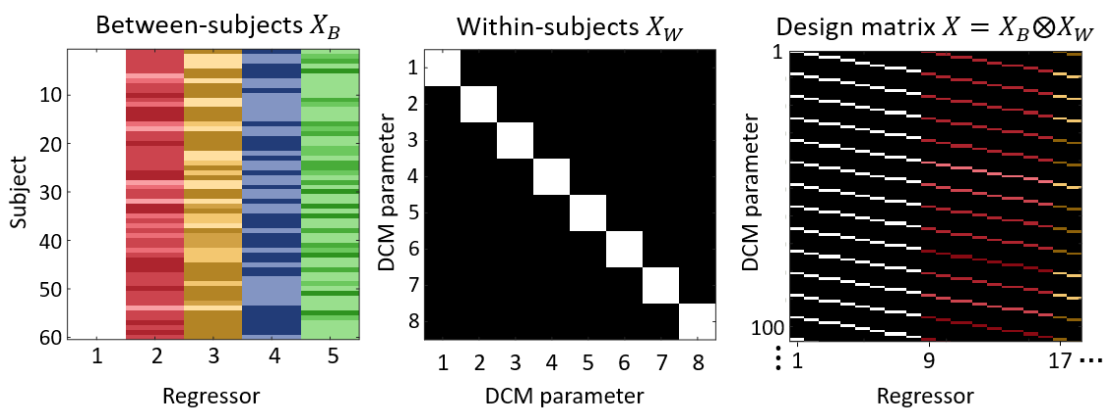

**Figure 3 The group-level design matrix X**. **Left**: the between-subjects design matrix, with regressors (covariates) modelling the group mean, laterality index, handedness, gender and age.

**Middle**: the within-subjects design matrix, where the diagonal encodes which DCM parameters receive group-level effects. This was set to the identity matrix to include all between-subject effects on all DCM parameters. **Right**: The resulting design matrix, computed according to Equation 5. Each row corresponds to one DCM parameter from one subject and each column corresponds to one group-level effect (e.g. gender) on one DCM parameter. For display purposes only, the regressors were z-scored and thresholded to produce colours in a consistent range.

## 5.3 PEB: Model estimation

Having specified the design matrix one can now invert the PEB model, which furnishes two useful quantities: the estimated group-level parameters and the group-level free energy. The parameter estimates correspond to a multivariate Gaussian density:

$$p(\boldsymbol{\theta}^{(2)}|Y) \sim N(\boldsymbol{m}_\theta^{(2)}, \boldsymbol{S}_\theta^{(2)}) \qquad 6.$$

Where vector $\boldsymbol{m}_\theta^{(2)}$ are the estimated betas or weights on the covariates in the design matrix $X$ and the covariance matrix $\boldsymbol{S}_\theta^{(2)}$ specifies uncertainty over these weights. (This should not be confused with $\boldsymbol{\Sigma}^{(2)}$ in Equation 3, which specifies the between-subject variability.)

Figure 4 (left) shows the estimated second level parameters pertaining to the effects of interest; namely, the group average and the effect of LI (**spm_dcm_peb_review.m**). The heights of the bars are $\boldsymbol{m}_\theta^{(2)}$ in Equation 6 and the error bars were computed from the leading diagonal of covariance matrix $\boldsymbol{S}_\theta^{(2)}$. Parameters 1-8 are the commonalities (group average connectivity parameters) across all subjects and parameters 9-16 are the differences in connectivity due to LI score. Each parameter quantifies the change in an inhibitory self-connection due to words or pictures.

To focus on a specific example – which will be relevant to the analyses which follow – consider the DCM parameter quantifying the effect of words on the inhibitory self-connection of region rdF. This was represented by two PEB GLM parameters, indicated in Figure 4 (left) with dashed lines. Parameter number eight is the average effect of words on rdF across subjects and parameter number 16 is the difference in this effect between subjects due to LI. From this plot, we can see that words increased region rdF's inhibitory self-connection. The effect size was 0.41 in the units of the underlying DCM. Parameter number 16 shows that the effect of words on rdF also correlated positively with LI score. The effect size was 1.79, meaning that the contribution of LI was to increase the rdF self-connection by 1.79 times the LI score. Note that these are just unstandardized effects sizes; to formally test hypotheses about where laterality effects are manifest, model comparisons are required, which we will return to shortly. Figure 4 (right) shows the estimated between-subject variability (i.e., random effects) of the eight DCM connectivity parameters, corresponding to the leading diagonal of matrix $\boldsymbol{\Sigma}^{(2)}$ in Equation 3.

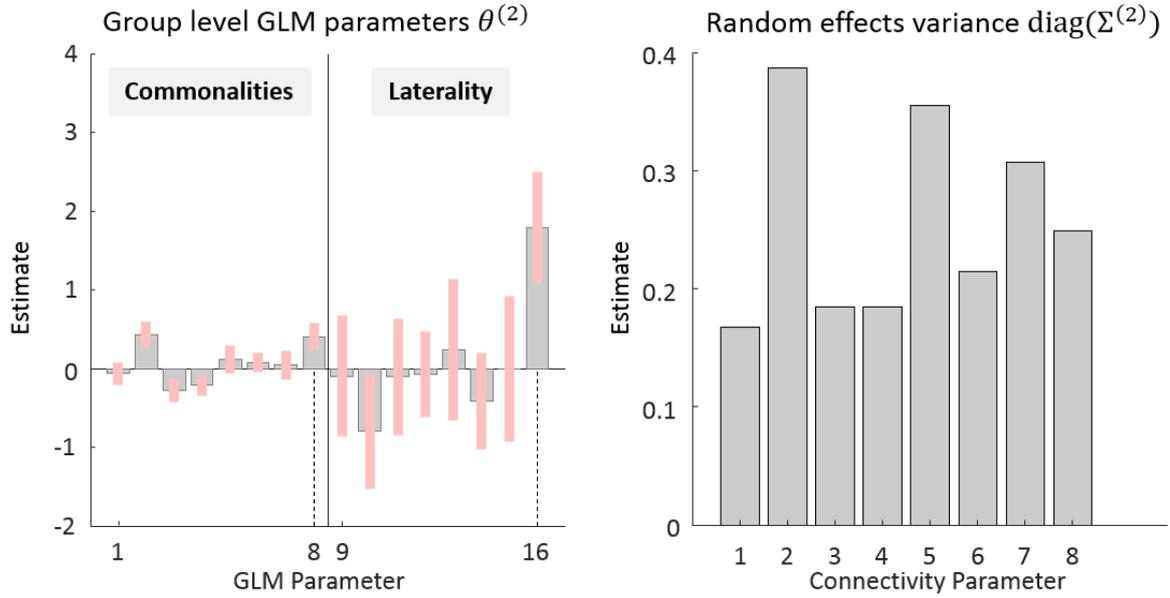

**Figure 4 Parameters of the group-level Bayesian General Linear Model (GLM). Left:** the posterior parameter estimates, corresponding to the first 16 regressors in design matrix X. These represent the commonalities (average) across subjects (GLM parameters 1-8) and the difference in connectivity due to Laterality Index (LI, GLM parameters 9-16). The units are the same as the parameters in the underlying DCMs; i.e., unitless log scaling parameters that scale the default self-connection strength of -0.5Hz. Dotted lines indicate parameters relating to the modulation of region rdF by words, discussed in the text. **Right**: the estimated between-subject variance of each connectivity parameter, after accounting for known effects. Parameters have the same order in both plots: 1=Modulation (B) of lvF by pictures, 2=ldF by pictures 3=rvF by pictures, 4=rdF by pictures, 5=lvF by words, 6=ldF by words, 7=rvF by words, 8=rdF by words.

Inversion of the full PEB model also returned a score for its quality – the (negative) variational free energy $F^{(2)}$ – which approximates the log model evidence:

$$F^{(2)} \approx \ln p(Y|m) \qquad 7.$$

This is the log of the probability of observing the neuroimaging data (from all subjects) given the entire hierarchical model $m$. Note there is a distinction between the free energy at the first level (relating to individual subjects' DCMs) and the second level (relating to the entire group and our model of between subject effects). At the first level, the free energy of a DCM is its accuracy minus its complexity; where complexity is the difference (KL-divergence) between the estimated parameters and the priors. At the second level, the free energy is the sum of all subjects' DCMs accuracies, minus the complexity induced by fitting the DCMs *and the second-level GLM* (see Equation 10 of Friston et al. (2016)). By comparing the free energy of PEB models – with different sets of GLM parameters switched on and off – and selecting the PEB models with the greatest free energy, one can now find the optimal explanation for the dataset as a whole.

## 5.4 Inference: Comparing reduced PEB models

Having estimated the parameters $\boldsymbol{\theta}^{(2)}$ of the group-level GLM, we next test hypotheses, finding out whether there is an effect of laterality and, if so, where it is expressed. In the PEB framework, this is done by comparing the evidence for reduced GLMs that have certain combinations of parameters 'switched off' (fixed at their prior expectation of zero). Comparing full and reduced models in this way is conceptually similar to performing an F-test in classical statistics. In practice, the evidence and parameters of the reduced models can be derived analytically from the full model in a matter of milliseconds, using a procedure referred to as Bayesian Model Reduction (BMR, see *Appendix 3: Bayesian Model Comparison and Reduction*).

We defined a set of candidate models to identify the best explanation for the commonalities across subjects and the best explanation for the LI differences, in terms of the three experimental factors defined above (Figure 2). Each factor had three levels, plus a null hypothesis, necessitating $3 \times 3 \times 3 + 1 = 28$ candidate models to evaluate all combinations of the factors, shown schematically in Figure 5A. This is referred to as a factorial model space, because each of the three factors can be imagined as an axis on a graph, with each of the models situated somewhere in the space formed by those axes.

We asked which of the 28 models (patterns of switched on / switched off parameters) was the best explanation for the commonalities across subjects and which was the best explanation for the LI differences. This meant comparing the evidence for a total of $28 \times 28 = 784$ reduced GLMs (**spm_dcm_peb_bmc.m**). Figure 5B shows the resulting matrix $\boldsymbol{P}$ of posterior probabilities, where element $P_{i,j}$ is the probability of the GLM that had its commonalities parameters configured according to model $i$ and its LI parameters configured according to model $j$. It is clear from Figure 5B that one GLM stood out (56% posterior probability). Its commonalities parameters were deployed according to model 4 and its LI parameters were deployed according to model 15. This result is shown more clearly by marginalising (summing) over the columns and rows and re-normalising, to give the probability for each model of the commonalities (Figure 5C) and LI differences (Figure 5D). Model 4, which was the best explanation for the commonalities across subjects, had modulation by words and pictures in dorsal regions (Figure 5E). Model 15, which was the best explanation for the differences across subjects due to LI, had words specifically modulating the self-connection of rdF (Figure 5F).

Despite the relatively strong probability for these two models, no single model could be described as an overall winner (> 95% probability), which is unsurprising given the large number of models. To summarise the parameters across all models, we computed the Bayesian Model Average (BMA). This is the average of the parameters from different models, weighted by the models' posterior probabilities (Hoeting et al., 1999; Penny et al., 2006). The averaged parameters are shown in Figure 5G and the values are listed in Table 2. To focus our analysis on just those parameters that had the greatest evidence of being non-zero, we thresholded the BMA to only include parameters that had a 95% posterior probability of being present vs absent (this thresholding based on the free energy is detailed in Appendix 3). As illustrated in Figure 5H, in common across subjects, pictures increased self-inhibition in region ldF and dis-inhibited rdF; shifting the balance of activation towards the right hemisphere with no differences across subjects due to LI score. By contrast, words increased self-inhibition in rdF, which was further increased in subjects with more positive LI scores; i.e. with greater left-lateralisation of brain responses. We may conclude, therefore, that individual differences in LI score could be explained by differences in the gain or excitability of region rdF, when presented with words.

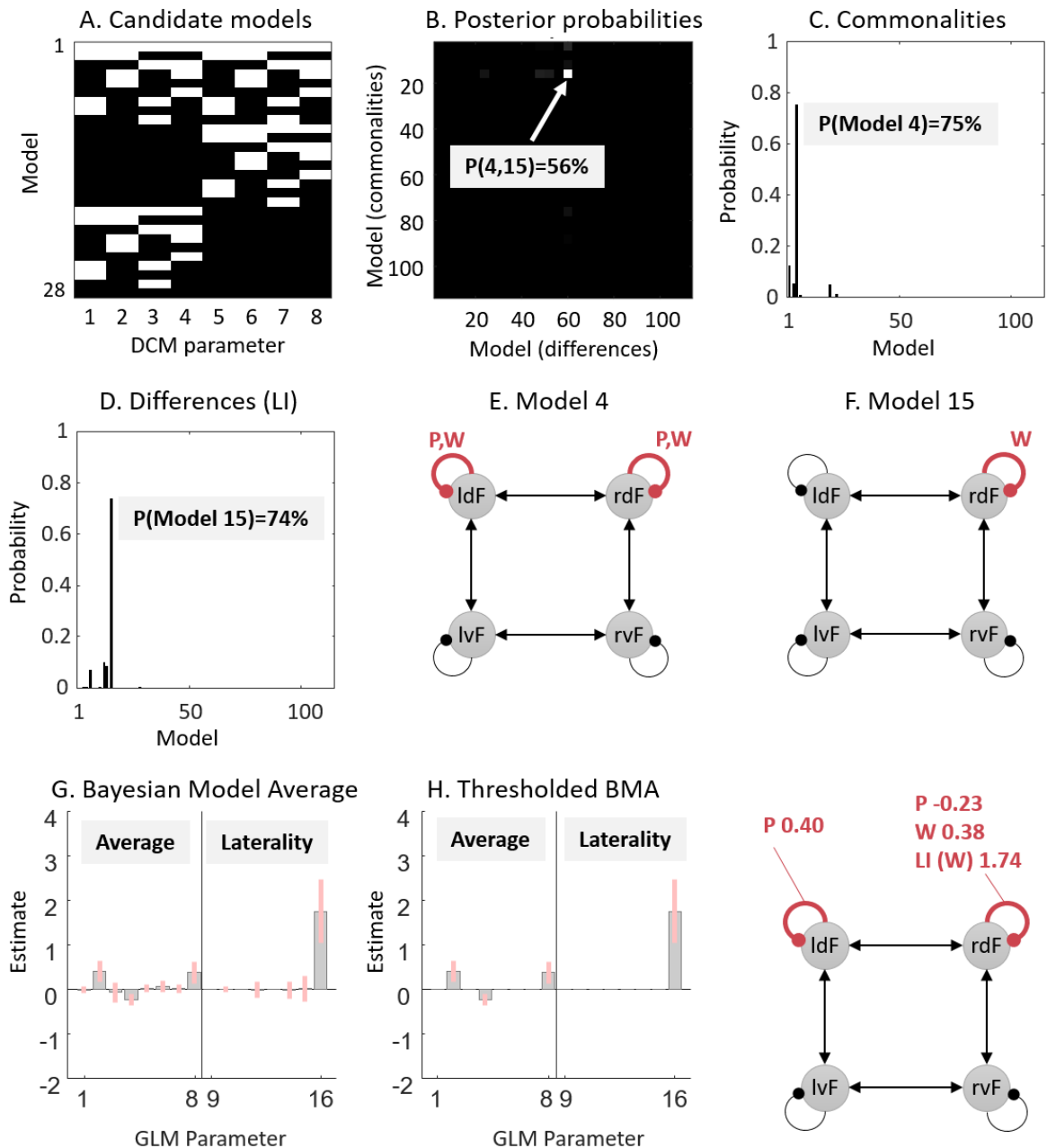

**Figure 5** Comparison in a pre-defined model space. **A.** Connectivity parameters switched on (white) and off (black) in each model. See the legend of Figure 4 for the identity of each parameter. **B.** Joint probability of all candidate models. The model in row $i$ column $j$ had its commonalities parameters set according to model $i$ and its laterality index (LI) parameters set according to model $j$. The best model was number 4 for the commonalities and 69 for laterality differences, with 56% posterior probability. **C.** The same result shown in part B, summed over the columns and re-normalized, to give the posterior probability for the commonalities across subjects. **D.** The same result shown in panel B, summed over the rows and re-normalized, to give the posterior probability for the models of laterality index (LI). **E-F.** Schematic diagrams of models 4 and 15. Curved lines are self-connections modulated by words and / or pictures. **G.** Bayesian Model

Average (BMA) of the parameters over all models. **H.** The BMA thresholded at posterior probability > 95% for clarity. The schematic illustrates the parameters which survived thresholding. P=Pictures, W=Words, LI=Laterality Index

## 5.5 Inference: family analysis

In the model comparison above, there were three factors with three levels each, plus a 'null' model, giving $3 \times 3 \times 3 + 1 = 28$ models of the commonalities across subjects and 28 models of the LI differences, thus $28 \times 28 = 784$ candidate models in total. Rather than conclude the analysis by plotting 784 values, it is clearer to provide a single statistic for each of the three factors. To do this, we grouped the 784 models into 'families' according to each of the experimental factors and compared the pooled evidence of each family (Penny et al., 2010). The results, computed using **spm_dcm_peb_bmc_fam.m**, are shown in Figure 6 and are described in the sections which follow.

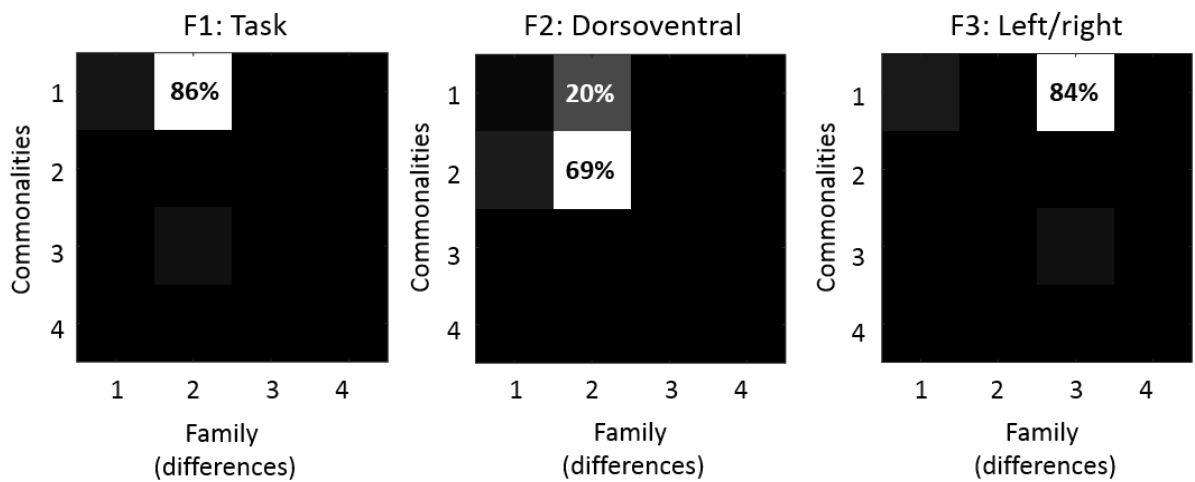

**Figure 6 Family-wise analyses for each of three experimental factors (F1-3).** Each plot is a separate analysis with the same 784 PEB models grouped into different families. The families are illustrated in Figure 2 (additionally, family 4 was added to each analysis, containing just the null model with no modulation by words or pictures). For each plot, element $P(m,n)$ represents the pooled probability for models in which the commonalities parameters were set according to family $m$ and the Laterality Index parameters were set according to family $n$. For example, for Factor 1, element $P(1,2)$ is the pooled probability for all models with commonalities parameters set according to family 1 (pictures and words modulating connections) and Laterality Index parameters set according to family 2 (only words modulating connections).

### 5.5.1 Factor 1: Was the network modulated by whether the stimuli were words and / or pictures?

We grouped the 784 GLMs into four families according to the Task factor (see Figure 2), plus the null model with no modulation. For example, family 1 included all models that had both pictures and words modulating the network. The evidence for the models within each family was pooled, and the families were compared under the prior that each family was equally likely. There was 86% probability that model 1 (pictures and words both modulating) was the best explanation for the commonalities and that model 2 (only words) was the best model for LI differences. Therefore, we found that the frontal network was modulated by words and pictures, but LI differences were specifically explained by the response to words.

### 5.5.2 Factor 2: Did the stimuli modulate dorsal and / or ventral regions?

We next grouped the same 784 models into families defined by factor 2 (dorsal vs ventral). The probability that family 2 (dorsal regions) best explained both the commonalties and the laterality differences was 69%. The second best model, with 20% probability, had family 1 (both dorsal and ventral regions) for the commonalities. We can conclude that task effects were mainly expressed in the dorsal regions, and the effect of LI was specific to dorsal regions.

### 5.5.3 Factor 3: Did the stimuli modulate left and / or right hemisphere?

Finally, we organised the same models according to factor 3. There was 84% probability that family 1 (both left and right regions) were modulated by words and pictures, whereas family 3 (right hemisphere) was the best explanation for differences due to LI. Therefore, differences in hemispheric asymmetry across subjects could be explained specifically by differences in the right hemisphere.

## 5.6 Interim summary: Hypothesis-based analyses

To summarise the analyses so far, we first estimated a single DCM for each subject with all DCM parameters of interest switched on. We then formed a 'full' second level GLM with parameters representing the group average connection strengths (commonalities), the differences between subjects due to LI, and covariates of no interest (handedness, age and gender). We then compared 784 reduced versions of this GLM with different combinations of parameters – representing the commonalities and laterality differences across subjects – switched off. This provided free energy scores quantifying the evidence for each candidate model. By creating families of reduced GLMs – and pooling their evidence – we were able to address the effect of between-subject factors on within-subject parameters (i.e., condition specific changes in connectivity).

## 5.7 Inference: Search over reduced PEB models

In the analyses above, large numbers of pre-defined models were compared (in seconds) using a technology called Bayesian Model Reduction (BMR). If there were no strong hypotheses about between-subject effects on connectivity, and the objective was simply to 'prune' any GLM parameters that did not contribute to the model evidence, BMR can be used to automatically search over reduced models. This more exploratory approach is conducted under the assumption that all reduced models are equally probable *a priori*, and thus the 'full' model only contains parameters that are biologically plausible.

For this experiment, the GLM contained 40 parameters in total (*five* between subject effects on *eight* DCM parameters – Figure 3, right) and the objective was to find the best reduced GLM with certain parameters switched off. Evaluating all possible reduced GLMs would require evaluating $2^{40} = 1.099 \times 10^{12}$ models, which is not possible in a reasonable amount of time. Therefore, to reduce the computational demand, an automatic search procedure is used (**spm_dcm_peb_bmc.m**), originally referred to as 'post-hoc' DCM analysis (Friston and Penny, 2011; Rosa et al., 2012). This procedure compares the evidence for reduced models using Bayesian Model Reduction, iteratively discarding parameters that do not contribute to model evidence. The iterative procedure stops when discarding any parameter starts to decrease model evidence. Technically, this is known as a greedy search and allows thousands of models to be compared quickly and efficiently.

A Bayesian Model Average (BMA) is then calculated over the 256 models from the final iteration of the greedy search, and the result for our data is shown in Figure 7. The middle and right plots show the parameters with a posterior probability of being non-zero greater than 0.95. It is clear that the estimated response to words and the difference in the response to words due to LI were remarkably similar to the analysis above; however, in this illustrative analysis we did not have to declare any specific *a priori* models. The estimated response to pictures was slightly different, effecting rvF rather than rdF. If this difference was of experimental interest, these two explanations could be compared in terms of their free energy and the relative probability of each explanation quantified.

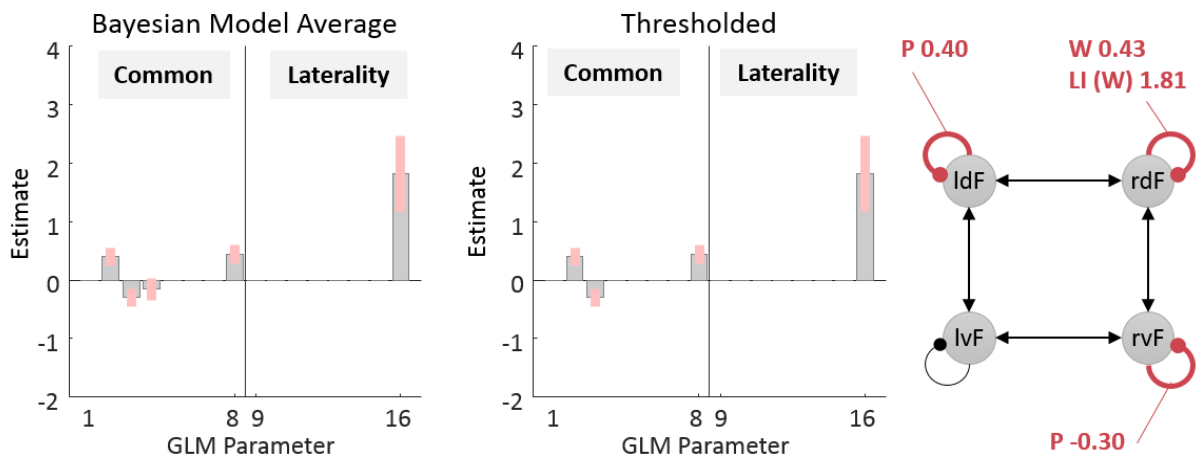

**Figure 7 BMA performed on the final 256 models of an automatic parameter search. Left:** Posterior parameter estimates based on the BMA. Only parameters relating to commonalities and Laterality Index are shown. **Middle**: Thresholded BMA, where parameters with probability greater than 95% of being present vs. absent were retained. Each parameter's individual probability was assessed by comparing the evidence for all models (from the final iteration of a greedy search) which had the parameter switched on, versus all models which had the parameter switched off. **Right**: The schematic shows effects of Pictures (P), Words (W) and Laterality Index (LI).

## 5.8 Inference: Network structure

Although our hypotheses were about changes in effective connectivity due to words and pictures (parameter matrix $B$ from the DCM neural model), it is useful to present these results in the context of the average effective connectivity across experimental conditions (parameter matrix $A$). This is because the $B$ parameters encode the deviation from the average connection strengths $A$ for each experimental condition. To bring these parameters to the group level, we specified and estimated a separate PEB model for the average connectivity $A$, and performed an automatic search over reduced models. While these analyses could have been done in the same PEB model, including a large number of parameters can cause a dilution of evidence effect, as well as inducing a much larger search space. Examining each set of parameters separately can therefore help to keep the analysis focused and tractable. The parameter estimates are listed in Table 3. There were two non-trivial effects of Laterality Index: lvF->ldF (-0.35Hz) and rdF->ldF (0.48Hz). Thus, in subjects with greater left hemisphere dominance as scored by a more positive LI, region ldF was more inhibited by lVF and more excited by rdF, independent of whether the stimuli were pictures or words.

# 6  Prediction: Cross-validation

In the previous steps, we identified a laterality effect on the response of region rdF to words. Our final question was whether the size of this effect was large enough to be interesting; i.e., whether we could predict a subject's LI from their neural response. Questions of this sort – assessing *predictive validity* – are particularly important for studies determining the clinical significance of model parameters. To address this we used leave-one-out cross validation (**spm_dcm_loo.m**). A PEB model was fitted to all but one subject, and covariates for the left-out subject were predicted. This was repeated with each subject left out and the accuracy of the prediction was recorded. For the technical details of this procedure see *Appendix 4: Leave-one out cross-validation with PEB*.

We assessed whether we could predict subjects' LI based on their modulation of rdF by words, as well as their known covariates - handedness, age and gender. The red line in Figure 8 (left) shows the predicted (mean-centred) LI score for each left out subject. The shaded area is the 90% credible interval of the prediction and the dashed black line is the actual mean-centred LI. In this example, 44 out of 60 of subjects had their true LI within the estimated 90% credible interval (shaded area). Figure 8 (right) plots the out-of-samples correlation of the actual LI against the (expected value of) the predicted LI for each left-out subject. The Pearson's correlation coefficient was 0.34, p=0.004. Therefore, we can conclude that the effect size estimated by DCM was sufficiently large to predict the left-out subjects' LI with performance above chance, although there was still a lot of variability to be explained.

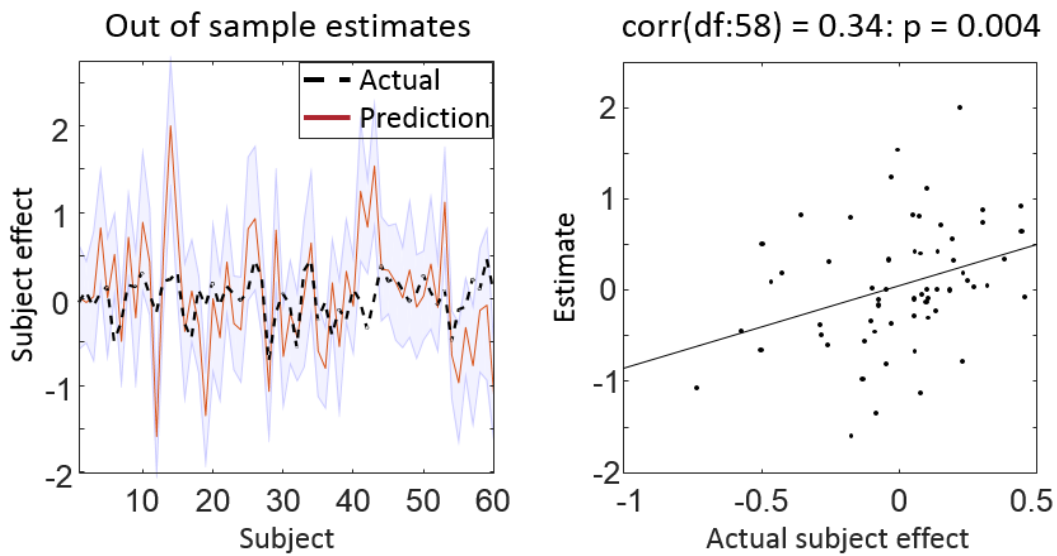

**Figure 8 Leave-one-out cross validation. Left**: the out-of-samples estimate of (mean-centred) Laterality Index for each subject (red line) with 95% confidence interval (shaded area). The black dashed line is the actual group effect. **Right**: The actual subject effect plotted against the expected value of the estimated subject effect.

# 7 Analysis summary

The above analyses characterised the effective connectivity underlying individual differences in lateralisation, in the context of a semantic processing task. We asked whether the commonalities and LI differences across subjects were expressed in condition specific changes in connectivity due to picture and / or word stimuli, in dorsal and / or ventral regions, and in left and / or right hemispheres (Figure 2). Comparisons of PEB models showed that in common across subjects, the frontal network was modulated by pictures and words, in dorsal regions of both hemispheres (Figure 6). The parameters revealed a double dissociation between picture and word stimuli: pictures inhibited responses in region ldF, whereas words inhibited responses in rdF (Figure 5H). Differences between subjects – due to LI score – were specifically expressed in the response to words in the dorsal region of the right hemisphere (Figure 5H and Figure 6). This effect was positive, meaning that having a higher LI score (being more left lateralized) was associated with greater inhibition of region rdF. Cross-validation showed that this effect was sufficiently large to predict left-out subjects' LI.

# 8 Discussion

We have illustrated an empirical (i.e. hierarchical) Bayesian procedure for conducting group connectivity analyses, using DCM and PEB. This begins with a first level analysis, modelling within-subject effects using neural models (DCMs). The connectivity parameters of interest are then taken to a second level analysis, where they are modelled using a Bayesian GLM. Hypotheses are tested by comparing the evidence for the full GLM against reduced GLMs, where certain parameters are switched off (i.e., fixed at zero). Finally, the predictive validity of the model parameters is assessed using cross-validation.

The PEB framework (Friston et al., 2016) introduced hierarchical modelling of neural parameters to DCM, linking individual subjects to the group level. This has so far proved useful for modelling fMRI data (Dijkstra et al., 2017; Park et al., 2017; Jung et al., 2018; Zhou et al., 2018a; Zhou et al., 2018b) and MEG / EEG / LFP data (Pinotsis et al., 2016; Fardo et al., 2017; Papadopoulou et al., 2017; Penny et al., 2018). It has been validated in terms of reproducibility in the context of ERP data (Litvak et al., 2015) and – going forward – will benefit from further validation to assess the suitability of certain priors in novel applications. While this study has demonstrated the main features of the framework, there are others we did not cover for brevity. For instance, to pull subjects' parameter estimates out of local minima, the first level model estimation can be re-initialized multiple times, using the group-level posteriors from the PEB model as *empirical priors* for each subject (Friston et al., 2015). This may be particularly useful where highly non-linear DCMs are in play. Furthermore, PEB analyses do not need to be limited to two levels. In a recent study investigating changes in the motor system following thalamic surgery for tremor, resting state fMRI data from multiple time points were modelled using separate DCMs (Park et al., 2017). The DCM parameters from each time point were collated into a subject-specific PEB model, forming the second level of the hierarchy. The second level parameters were then taken up to the group level, modelled using a third-level GLM, to capture commonalities and differences among subjects. By using this PEB-of-PEBs approach, the deep hierarchies implicit in certain kinds of experimental designs can be modelled explicitly.

Here, we presented two methods for testing hypotheses at the group level – either designing a set of models / families explicitly, or automatically searching over potentially thousands of reduced models to 'prune' parameters that did not contribute to model evidence. While the two approaches gave similar results on the dataset presented here, this is not guaranteed to be the case generally, and there are disadvantages to using an automatic (greedy) search. The large number of possible models means the search cannot usually be exhaustive, and so a better solution may be found by a well thought-out set of models. Furthermore, an automatic model search introduces the temptation to construct post-hoc explanations for the surviving parameters. Given the number of possible reduced models, one might emphasise the utility of *a priori* hypotheses when using Bayesian model reduction.

Specifying the PEB model in this study involved setting the between-subjects design matrix *a priori*. A complementary application of hierarchical modelling would be to discover group membership; i.e., to perform unsupervised classification or clustering of subjects based on their connectivity. A recently developed approach for this combines DCM with finite mixture models in a single hierarchical model (Raman et al., 2016). A complementary approach is possible with the PEB scheme, by searching over candidate group assignments using Bayesian Model Reduction (BMR). It would be interesting to compare this against sampling approaches, which are generally slower but constitute the gold standard in terms of robustness.

Future studies will be able to evaluate other advantages of the PEB approach. For instance, we would expect parameters from outlier subjects to be suppressed by the constraints placed upon them by group-level empirical priors. Additionally, estimating a single 'full' DCM for each subject may offer additional benefit: we speculate that a 'full' model with all parameters in play will typically engender a 'smoother' free energy landscape than reduced models with fewer parameters. In other words, having a high dimensional parameter space can, counterintuitively, ensure more robust convergence; in the sense that there are more opportunities to escape from local minima in high dimensional parameter spaces. This may be particularly prescient for electrophysiological DCMs, where the models are more non-linear. To get started with the PEB framework, we refer readers to the illustrated step-by-step guide and example data accompanying this article at https://github.com/pzeidman/dcm-peb-example.

# 9 Tables

Table 1: Symbols used in this paper

| Variable | Dimension | Meaning |
| --- | --- | --- |
| $\alpha$ | $1 \times 1$ | Ratio of prior uncertainty about group-level effects to prior uncertainty about first-level effects. |
| $\beta$ | $1 \times 1$ | Ratio of within-subject prior variance (uncertainty) to between-subject prior RFX variance (variability). |

| Symbol | Dimensions | Description |
| --- | --- | --- |
| $\beta_0$ | $C_0 \times 1$ | Parameters for null effects |
| $C$ | $1 \times 1$ | Total group level covariates |
| $C_0$ | $1 \times 1$ | Total first level covariates of no interest |
| $\epsilon_i^{(1)}$ | $V \times 1$ | Observation noise for subject $i$ |
| $\epsilon^{(2)}$ | $(N \cdot P^{(1)}) \times 1$ | Between-subject variability or random effects (RFX) |
| $\epsilon^{(3)}$ | $(C \cdot P^{(1)}) \times 1$ | Residuals (uncertainty) about group-level parameters $\theta^{(2)}$ |
| $\eta$ | $(C \cdot P^{(1)}) \times 1$ | Expected values of group-level parameters $\theta^{(2)}$ |
| $F^{(n)}$ | $1 \times 1$ | Negative variational free energy for level $n$ of the model |
| $\Gamma_i$ | — | First-level model (e.g. DCM) for subject $i$ |
| $J$ | $1 \times 1$ | Number of experimental conditions at the first level |
| $\gamma_j$ | $1 \times 1$ | Log scaling parameter for precision component $Q_j$ |
| $L$ | $1 \times C$ | Norm of each covariate (regressor) |
| $m$ | — | Generative model (likelihood and priors) |
| $\mu_i, \mu_i^{(1)}$ | $P^{(1)} \times 1$ | Expected values of the DCM parameters for subject $i$ |
| $\mu_0^{(1)}$ | $P^{(1)} \times 1$ | Prior expectation of the DCM parameters |
| $N$ | $1 \times 1$ | Number of subjects (not to be confused with the multivariate normal distribution $N(\mu, \Sigma)$). |
| $P^{(1)}$ | $1 \times 1$ | Total free parameters per DCM taken to the group level |
| $\Pi^{(2)}$ | $(C \cdot P^{(1)}) \times (C \cdot P^{(1)})$ | RFX precision matrix |
| $Q_j$ | $P^{(1)} \times P^{(1)}$ | Precision component $j = 0 \ldots J$ |
| $R$ | $1 \times 1$ | Number of modelled brain regions |
| $\Sigma_i, \Sigma_i^{(1)}$ | $P^{(1)} \times P^{(1)}$ | Covariance matrix of the DCM parameters for subject $i$ |
| $\Sigma_0^{(1)}$ | $P^{(1)} \times 1$ | Prior covariance of the DCM parameters |
| $\theta^{(1)}$ | $NP^{(1)} \times 1$ | All DCM parameters from all subjects |
| $\theta_i^{(1)}$ | $P^{(1)} \times 1$ | All DCM parameters for subject $i$ |
| $\Sigma^{(2)}$ | $(N \cdot P^{(1)}) \times (N \cdot P^{(1)})$ | Covariance of between-subject variability $\epsilon^{(2)}$ |
| $\Sigma^{(3)}$ | $(C \cdot P^{(1)}) \times (C \cdot P^{(1)})$ | Covariance of residuals (uncertainty) $\epsilon^{(3)}$ |
| $\theta^{(2)}$ | $(C \cdot P^{(1)}) \times 1$ | Group-level parameters |
| $V$ | $1 \times 1$ | Total measurements (volumes) per subject |
| $v_j$ | $1 \times 1$ | Prior variance of DCM parameter $j$ |
| $X$ | $(N \cdot P^{(1)}) \times (C \cdot P^{(1)})$ | Design matrix |
| $X_B$ | $N \times C$ | Between-subjects design matrix |
| $X_W$ | $P^{(1)} \times P^{(1)}$ | Within-subjects design matrix |
| $X_0$ | $V \times C_0$ | Design matrix for null effects |
| $Y_i$ | $V \times R$ | Observed timeseries from subject $i$ from all regions |

Table 2: BMA of PEB parameters: Expected values of estimated commonalities and LI differences

|  | Commonalities | | Laterality Index | |
| --- | --- | --- | --- | --- |
| Region | Pictures | Words | Pictures | Words |
| lvF | -0.02 | 0.02 | 0 | 0 |
| ldF | 0.40 | 0.06 | -0.001 | -0.02 |
| rvF | -0.07 | 0 | 0 | 0.01 |
| rdF | -0.23 | 0.38 | -0.02 | 1.74 |

* 'Pictures' and 'Words' are the log of scaling parameters that multiply up or down the default self-connection (-0.5Hz).

Table 3: BMA of PEB parameters: Average connectivity across experimental conditions (A-matrix)

| Source | Target | Commonalities | Laterality index |
| --- | --- | --- | --- |
| lvF | lvF | -0.20 | 0 |
| lvF | ldF | 0.16 | -0.35 |
| lvF | rvF | 0.08 | 0 |
| ldF | lvF | 0.08 | 0 |
| ldF | ldF | -0.18 | 0 |
| ldF | rdF | 0.06 | 0 |
| rvF | lvF | 0.10 | 0 |
| rvF | rvF | -0.30 | 0 |
| rvF | rdF | 0.08 | 0 |
| rdF | ldF | 0.18 | 0.46 |
| rdF | rvF | 0 | 0 |
| rdF | rdF | -0.32 | 0 |

* Between-region connections are in units of Hz. Self-connections, where the source and target are the same, are the log of scaling parameters that multiply up or down the default value -0.5Hz.

# 10 Appendix 1: PEB connectivity priors specification

In the PEB framework, the first-level models have priors defined by a multivariate normal density $p(\boldsymbol{\theta}^{(1)}) = N(\boldsymbol{\mu}_0^{(1)}, \boldsymbol{\Sigma}_0^{(1)})$, which are the same for each subject. The second-level model is a Bayesian GLM with parameters $\boldsymbol{\theta}^{(2)}$. For convenience, the priors on these parameters $p(\boldsymbol{\theta}^{(2)})$ are derived from the first level priors. By default, the second level priors are identical to those of the first level, except the prior variance is adjusted based on the scaling of the design matrix $X$. The norm of the columns of between-subjects design matrix $X_B$ is calculated as follows:

$$L = \frac{N}{\text{SS}(X_B)} \qquad 8.$$

Where function SS is the sum-of-squares and $N$ is the number of subjects. The second level prior is defined according to:

$$p(\boldsymbol{\theta}^{(2)}) = N(\boldsymbol{\eta}, \boldsymbol{\Sigma}^{(3)})$$

$$\boldsymbol{\eta} = \boldsymbol{\mu}_0^{(1)}$$

$$\boldsymbol{\Sigma}^{(3)} = \frac{L \otimes \boldsymbol{\Sigma}_0^{(1)}}{\alpha}$$

9.

This specification has the effect of scaling the prior variance of all second level parameters relating to a particular covariate by the norm of that covariate. Here, we set the fixed parameter $\alpha$ to one, meaning that we expected the effect sizes at the second level to be the same as the first level (aside from the scaling).

## 11 Appendix 2: PEB random effects specification

Any differences between subjects not captured by the second-level Bayesian GLM are defined as additive noise or random effects (RFX) $\boldsymbol{\epsilon}^{(2)}$:

$$\boldsymbol{\epsilon}^{(2)} \sim N(\mathbf{0}, \boldsymbol{\Sigma}^{(2)})$$

$$\boldsymbol{\Pi}^{(2)} = \boldsymbol{\Sigma}^{(2)^{-1}} = \boldsymbol{I}_S \otimes \left( \boldsymbol{Q}_0 + \sum_j e^{-\gamma_j} \boldsymbol{Q}_j \right)$$

10.

This is zero-mean additive noise with precision matrix (inverse of the covariance matrix) $\boldsymbol{\Pi}^{(2)}$. The second line of Equation 10 specifies the precision matrix as a weighted combination of $j = 1 \ldots J$ precision components. The Kronecker product $\otimes$ replicates across the $S$ subjects and $\boldsymbol{I}_S$ is the identity matrix. Precision component matrix $\boldsymbol{Q}_0$ is the minimum allowable precision for each parameter and each log scaling parameter $\gamma_j$ multiplied up or down the corresponding precision component $\boldsymbol{Q}_j$. These parameters are estimated from the data with prior $p(\gamma_i) = N(0, 1/16)$. Taking their exponential guarantees that the precision is positive.

Specifying this model requires selecting the precision components $\boldsymbol{Q}_{1 \ldots J}$. The PEB software provides several choices:

- A single precision component
- One component per field; i.e., one component for the DCM A-matrix parameters, one for the DCM B-matrix parameters, etc.
- One component per DCM connectivity parameter
- Custom allocation of parameters to components

We selected the third option, one component per DCM connection – meaning that the between-subject variability for each connection was individually estimated. The form of these components is illustrated in Figure 9. Each component $j = 1 \ldots J$ is a matrix with a single non-zero entry, which for convenience is set as a function of the corresponding first level DCM parameter's variance:

$$(\mathbf{Q}_j)_{j,j} = \left(\frac{v_j}{\beta}\right)^{-1} = \frac{\beta}{v_j} \qquad 11.$$

Where $v_j$ is the prior variance of any single subject's DCM parameter $j$ and fixed parameter $\beta = 16$. This equation says that we expect the between-subject variance to be smaller than the within-subject effect size; i.e. the expected between-subject standard deviation is a quarter of the prior standard deviation of the corresponding parameter. This means that the default between-subject precision for each parameter encoded on the leading diagonal of $\mathbf{Q}$ is $16/1 = 16$.

**Figure 9 The form of the precision components.** Each component $\mathbf{Q}_j$ has a single non-zero entry at element $(j,j)$, which is set to 16. These are scaled by the corresponding log scaling parameters $\gamma_j$. See Equation 10 for further details.

## 12 Appendix 3: Bayesian Model Comparison and Reduction

Hypotheses are tested in PEB by comparing the evidence for different models, which is called Bayesian model comparison or selection (c.f. Penny, 2012). The relative evidence for model $m_1$ relative to model $m_2$ is given by the Bayes factor in favour of model 1:

$$BF_1 = \frac{p(y|m_1)}{p(y|m_2)} \qquad 12.$$

In practice, the model evidence $p(y|m)$ cannot be computed exactly. In the DCM / PEB framework it is approximated by the (negative) variational free energy, which to recapitulate Equation 7 is:

$$F \approx \ln p(y|m) \qquad 13.$$

Having computed the free energy for two models $F_1$ and $F_2$, it is usually more convenient to work with the log of the Bayes factor:

$$\ln BF_1 = F_1 - F_2 \qquad 14.$$

This follows because division becomes subtraction when the values are logs. Sometimes the log Bayes factor is reported directly, or alternatively it can be expressed as the posterior probability for one model over another. This is given by Bayes rule under equal priors for each model:

$$
\begin{aligned}
p(m_1|y) &= \frac{p(y|m_1)}{p(y)} \\
&= \frac{p(y|m_1)}{p(y|m_1) + p(y|m_2)} \\
&= \frac{1}{1 + \frac{p(y|m_2)}{p(y|m_1)}} \\
&= \frac{1}{1 + BF_2} \\
&= \frac{1}{1 + \exp(\ln BF_2)} \\
&= \frac{1}{1 + \exp(-\ln BF_1)}
\end{aligned}
\qquad 15.
$$

The final line is simply a sigmoid function of the log Bayes factor. For example, a Bayes factor of 20 in favour of model 1 (20 times greater evidence for model 1 than model 2) is the same as a log Bayes factor of $\ln 20 = 3$, which is equivalent to posterior probability $p(m_1|y) = 0.95$. This value is typically used as the threshold for 'strong evidence' in favour of model 1 (Kass and Raftery, 1995). This approach can easily be generalized to the case where there are more than two models, by computing the log Bayes factor of each model relative to the model with the lowest evidence.

Bayesian model comparison is predicated on computing the free energy for each model. In the setting where a 'full' model has been estimated, with all parameters of interest informed by the data, Bayesian Model Reduction (BMR) enables the free energy and parameters for 'reduced' models to be computed analytically under certain mild assumptions (Friston and Penny, 2011; Friston et al., 2016). The difference between a full and a reduced model is the priors: for example, in a reduced model, a certain subset of connections may be switched off (i.e., fixed at zero). This enables large numbers of models to be evaluated in seconds. BMR is used extensively in the PEB framework (**spm_log_evidence_reduce.m**), for example to search over potentially thousands of reduced models.

One application of BMR – used in this paper – was to threshold the parameters of a PEB model based on the free energy (e.g. Figure 5H). This involved, for each parameter $j$, performing a Bayesian model comparison of the PEB model with parameter $j$ 'switched on' (free to vary) versus the equivalent PEB model with parameter $j$ 'switched off' (fixed at its prior expectation of zero). The difference in evidence can then be converted to a posterior probability using Equation 15 – and it is common to focus discussion on the parameters that exceed a certain posterior probability threshold. The advantage of thresholding using this approach – rather than just selecting parameters that have marginal posteriors (error bars) that deviate from zero – is that the full covariance of the parameters is considered when computing a difference in free energy. A proper assessment is therefore made as to whether the parameter contributed to model evidence.

# 13 Appendix 4: Leave-one out cross-validation with PEB

The implementation of Leave-one out (LOO) cross-validation in the PEB framework (**spm_dcm_loo.m**) is most simply explained with an example. Consider a dataset with five subjects, each of whom has one parameter of interest in their first level DCM model. These subjects are drawn from two groups, where each group has mean parameter values 0.7 and 0.3. Following model estimation, their parameters have expected values: 0.7, 0.6, 0.8, 0.2 and 0.4 respectively. To create the training set, we leave out the first subject and specify a GLM on the remaining four:

$$\boldsymbol{\theta}^{(1)} = \boldsymbol{X}\boldsymbol{\theta}^{(2)} + \boldsymbol{\epsilon}^{(2)}$$

$$\begin{bmatrix} 0.6 \\ 0.8 \\ 0.2 \\ 0.4 \end{bmatrix} = \begin{bmatrix} 1 & 1 \\ 1 & 1 \\ 1 & -1 \\ 1 & -1 \end{bmatrix} \boldsymbol{\theta}^{(2)} + \boldsymbol{\epsilon}^{(2)} \qquad 16.$$

The design matrix $\boldsymbol{X}$ has two regressors – to model the group mean and the group difference. Calculating the parameters of this GLM, using the standard formula, gives:

$$\boldsymbol{\theta}^{(2)} = (\mathbf{X}^{\mathrm{T}}\mathbf{X})^{-1}\mathbf{X}^{\mathrm{T}}\boldsymbol{\theta}^{(1)} = \begin{bmatrix} 0.5 \\ 0.2 \end{bmatrix} \qquad 17.$$

This says that the group average connectivity is value 0.5, plus or minus 0.2 depending on the subject's group. Now we have an estimated model based on the training data, we can predict the row of the design matrix $\boldsymbol{X}_i$ for the left out subject $i$, given their connection strength $\boldsymbol{\theta}_i^{(1)}$. Specifically, we know the first element of this subject's row in the design matrix, but not the second:

$$\boldsymbol{X}_i = [1 \; x_i] \qquad 18.$$

Where $x_i$ is the unknown group membership. Recall that the left out subject had connectivity parameter $\boldsymbol{\theta}_i^{(1)} = 0.7$. We can construct a GLM for this subject as follows:

$$\theta_i^{(1)} = [\boldsymbol{\theta}^{(2)}]^T [\boldsymbol{X}_i]^T + \epsilon^{(2)} \Rightarrow$$

$$0.7 = [0.5 \; 0.2] \begin{bmatrix} 1 \\ x_i \end{bmatrix} + \epsilon^{(2)} \qquad 19.$$

Here, the role of the design matrix and parameters has been switched – we know the parameters $\boldsymbol{\theta}^{(2)}$, and we want to infer the missing element of the design matrix $x_i$. Solving this equation gives $x_i = 1$, which is the correct group membership assignment for this subject.

To apply this approach in the PEB framework, for predicting discrete or continuous variables, a group-level PEB model is fitted to the training data in the usual way, which includes all but the left-out subject $i$. The hierarchical model over parameters is therefore:

$$p\big(\boldsymbol{\theta}_{TRAIN}^{(1)} \big| \boldsymbol{\theta}_{TRAIN}^{(2)}\big) = N\big(X_{TRAIN} \cdot \boldsymbol{\theta}_{TRAIN}^{(2)}, \Sigma_{TRAIN}^{(2)}\big) \qquad 20.$$

Where subscript $TRAIN$ refers to training data. A separate PEB model is then fitted to the test data, which includes only the left out subject $i$, configured as follows:

$$p(\boldsymbol{\theta}^{(2)}_{TEST}) = N(\eta, \boldsymbol{\Sigma}^{(3)})$$

$$p(\boldsymbol{\theta}^{(1)}_{TEST}|\boldsymbol{\theta}^{(2)}_{TEST}) = N(\mathbf{X}_{TEST}\boldsymbol{\theta}^{(2)}_{TEST}, \boldsymbol{\Sigma}^{(2)}_{TEST})$$

$$\eta_p = \begin{cases} 0, & p = k \\ X_i(p), & p \neq k \end{cases}$$

$$[\boldsymbol{\Sigma}^{(3)}]_{m,m} = \begin{cases} 4 \cdot \mathrm{var}(X_{TRAIN}) + c, & m = k \\ c, & \text{otherwise} \end{cases}$$

$$\mathbf{X}_{TEST} = [\boldsymbol{\theta}^{(2)}_{TRAIN}]^T$$

21.

Where subscript $TEST$ indicates the test subject and $k$ is the index of the column in the design matrix we are trying to predict. The first two lines of Equation 22 recapitulate the definition of the second level model in the PEB framework. The third and fourth lines define the prior expectation and covariance respectively for the test subject's parameters. The prior expectation $\boldsymbol{\eta}$ is a vector where each element $\eta_p$ equals the corresponding element in the subject's row in the original design matrix $\boldsymbol{X_i}$, except for the value to be predicted, which is set to zero. The prior variance (uncertainty) for the parameter to be predicted is set to a multiple of the regressor's variance in the training data. To improve stability of the model estimation, a positive additive constant $c$ is added to the prior variance of all parameters, which decreases over repeated estimations of the left-out subject's model (with values $= [1, e^{-1}, e^{-2}, e^{-3}]$ ).